\documentclass[twocolumn]{aastex631}
\usepackage{natbib}
\graphicspath{{./}{figures/}}

\received{}
\revised{}
\revised{}
\accepted{}
\submitjournal{ApJ}

\shortauthors{Hasegawa et al.}

\begin{document}

\title{The appearance of a `fresh' surface on 596 Scheila as a consequence of the 2010 impact event}

\correspondingauthor{Sunao Hasegawa}
\email{hasehase@isas.jaxa.jp}

\author[0000-0001-6366-2608]{Sunao Hasegawa}
\affiliation{Institute of Space and Astronautical Science, Japan Aerospace Exploration Agency, 3-1-1 Yoshinodai, Chuo-ku, Sagamihara, Kanagawa 252-5210, Japan}

\author[0000-0001-8617-2425]{Micha\"{e}l Marsset}
\affiliation{Department of Earth, Atmospheric and Planetary Sciences, MIT, 77 Massachusetts Avenue, Cambridge, MA 02139, USA}
\affiliation{European Southern Observatory (ESO), Alonso de C\'{o}rdova 3107, 1900 Casilla Vitacura, Santiago, Chile}

\author[0000-0002-8397-4219]{Francesca E. DeMeo}
\affiliation{Department of Earth, Atmospheric and Planetary Sciences, MIT, 77 Massachusetts Avenue, Cambridge, MA 02139, USA}

\author[0000-0003-4191-6536]{Schelte J. Bus}
\affiliation{Institute for Astronomy, University of Hawaii, 2860 Woodlawn Drive, Honolulu, HI 96822-1839, USA}


\author[0000-0002-7332-2479]{Masateru Ishiguro}
\affiliation{Department of Physics and Astronomy, Seoul National University, Gwanak-gu, Seoul 08826, Republic of Korea}
\affiliation{SNU Astronomy Research Center, Seoul National University, Gwanak-gu, Seoul 08826, Republic of Korea}


\author[0000-0002-7363-187X]{Daisuke Kuroda}
\affiliation{Okayama Observatory, Kyoto University, 3037-5 Honjo, Kamogata-cho, Asakuchi, Okayama 719-0232, Japan}

\author[0000-0002-9995-7341]{Richard P. Binzel}
\affiliation{Department of Earth, Atmospheric and Planetary Sciences, MIT, 77 Massachusetts Avenue, Cambridge, MA 02139, USA}

\author[0000-0002-2934-3723]{Josef Hanu\v{s}}
\affiliation{Institute of Astronomy, Faculty of Mathematics and Physics, Charles University, V Hole\v{s}ovi\v{c}k\'{a}ch 2, 180 00 Prague 8, Czech Republic}

\author[0000-0001-6990-8496]{Akiko M. Nakamura}
\affiliation{Department of Planetology, Graduate School of Science, Kobe University, 1-1 Rokkodai-cho, Nada-ku, Kobe 657-8501, Japan}

\author[0000-0002-5033-9593]{Bin Yang}
\affiliation{European Southern Observatory (ESO), Alonso de C\'{o}rdova 3107, 1900 Casilla Vitacura, Santiago, Chile}
\affiliation{N\'{u}cleo de Astronom\'{i}a, Facultad de Ingenier\'{i}ay Ciencias, Universidad Diego Portales, Chile}

\author[0000-0002-2564-6743]{Pierre Vernazza}
\affiliation{Aix Marseille Univ, CNRS, LAM, Laboratoire d'Astrophysique de Marseille, Marseille, France}




\begin{abstract} 
Dust emission was detected on main-belt asteroid 596 Scheila in December 2010, and attributed to the collision of a few-tens-of-meters projectile on the surface of the asteroid. 
In such impact, the ejected material from the collided body is expected to mainly comes from its fresh, unweathered subsurface. 
Therefore, it is expected that the surface of 596 was partially or entirely refreshed during the 2010 impact. 
By combining spectra of 596 from the literature and our own observations, we show that the 2010 impact event resulted in a significant slope change in the near-infrared (0.8 to 2.5 $\mu$m) spectrum of the asteroid, from moderately red (T-type) before the impact to red (D-type) after the impact. 
This provides evidence that red carbonaceous asteroids become less red with time due to space weathering, in agreement with predictions derived from laboratory experiments on the primitive Tagish Lake meteorite, which is spectrally similar to 596. 
This discovery provides the very first telescopic confirmation of the expected weathering trend of asteroids spectrally analog to Tagish Lake and/or anhydrous chondritic porous interplanetary dust particles. 
Our results also suggest that the population of implanted objects from the outer solar system is much larger than previously estimated in the main-belt, but many of these objects are hidden below their space-weathered surface.
\end{abstract}



\section{Introduction} \label{sec:intro}
596 Scheila, a dark outer main-belt asteroid with a diameter of 114 km (mean value obtained from IRAS \citep{Tedesco2002}, AKARI \citep{Usui2011}, and WISE \citep{Masiero2011}) discovered in 1906, was found to exhibit a comet-like appearance on 2010 December 11, about 100 years after its discovery \citep{Larson2010CBET}.
596 is rather unique in that it is one of the only two main-belt asteroids larger than 100 km in diameter for which cometary-like activity was ever detected, and the second one being the dwarf-planet 1 Ceres \citep{Kuppers2014}.
All other main-belt comets previously discovered are less than 10 km in diameter \citep{Jewitt2012}.
Spectroscopic studies of 596 conducted after it became active appeared to suggest a distinct mechanism for dust emission compared to comets and other main-belt comets. 
No emission lines originating from volatile components were detected in its UV-visible spectrum from the Swift UV--optical telescope \citep{Bodewits2011}, the visible spectra by FORS2 onboard the Very Large Telescope \citep{Jehin2011CBET} and LRIS equipped Keck I \citep{Hsieh2012}, the near-infrared (NIR) spectrum by Spex on the InfraRed Telescope Facility (IRTF) \citep{Yang2011}, and the microwave spectrum by the Arecibo Observatory \citep {Howell2011IAUC}.
In addition, no coma derived from gases components was detected in the narrow-band images of 596 \citep{Jehin2011CBET}.
From spectroscopy in the 3 $\mu$m region, the surface of 596 was found to have no more than a few percent of water ice \citep{Yang2011}.
These observational results suggested that the cometary activity of 596 is not due to ice sublimation, although they do not themselves establish that the activity of 596 was not due to ice sublimation considering that the production rates of sublimation products could have been below the detection limits of the observations.

Based on analysis of the coma and dust tail distributions of 596, \cite{Jewitt2011,Moreno2011,Ishiguro2011a,Hsieh2012} confirmed that the cometary activity was not caused by ice sublimation, but by the impact of a projectile.
\citet{Ishiguro2011b} further succeeded in perfectly reproducing the observed dust tail morphology morphology of 596 from the model of ejecta distribution during an oblique impact observed in laboratory experiments. 
The authors were also able to determine the impactor size, impact timing and direction, the size of the resulting crater, and the tensile strength of 596's surface material.
The ejecta discharged from the impact crater was predicted to completely or partially cover the surface of 596 \citep{Ishiguro2011a,Bodewits2014}.
\citet{Licandro2011EPSC} obtained rotationally-resolved NIR spectra after the impact event, and showed that the post-impact spectra in the 0.8--2.4 $\mu$m wavelength region is homogeneous, meaning that the subsurface material ejected from the impact crater likely entirely resurfaced 596.

Our team is currently conducting a visible and NIR spectroscopic survey of main-belt asteroids with diameters larger than 100 km to investigate the physical properties of planetesimals that existed in the early solar system (see Introduction chapter in \citealt{Hasegawa2021b}).
In the course of compiling near-infrared data obtained in the past by the MIT--Hawaii Near-Earth Object Spectroscopic Survey (MITHNEOS)\footnote{http://smass.mit.edu/}, we found out that spectra of 596 from 0.8 $\mu$m to 2.5 $\mu$m wavelength region had been obtained before and after the 2010 impact as part of the survey.
So far, studies of the outcome of the impact on the physical properties of 596 have only been performed by means of visible lightcurve observations \citep[e.g.,][]{Ishiguro2011a,Bodewits2014}, and none of these studies focused on how the reflectance spectrum of 596 changed.
The purpose of this study is to investigate whether or not a change happened in the spectrum of 596 from 0.8 $\mu$m to 2.5 $\mu$m as a consequence of the impact event, as well as in the visible and 3-$\mu$m regions by studying literature. 
Then, we discuss the factors that may have contributed to this change.

\section{Observations and Data Analysis} \label{sec:Observations and Data Analysis}
NIR spectroscopic observations of 596 Scheila were conducted at two distinct epochs: 2002-06-01 and 2011-02-07, with the 3-meter IRTF located on Mauna Kea, Hawaii. 
Data from the first epoch of observation were published and analyzed by \citet{DeMeo2009} while data from the second epoch were never published. 
We describe here the observations and data reduction performed for the data acquired on 2011-02-07. 

We used the SpeX NIR spectrograph \citep{Rayner2003} combined with a 0.8$\times$15 arcsec slit in the low-resolution prism mode to measure the spectra over the 0.8--2.5 $\mu$m wavelength range. 
At the time of our observations, 596 had a visual V-band magnitude of 13.4 and was located at an airmass of 1.1. 
A total of 15 spectral images with 120\,s exposure time were recorded in an AB beam pattern to allow efficient removal of the sky background by subtracting pairs of AB images. 
The following calibration stars, known to be very close spectral analogs to the Sun, were observed on the same night: Hyades 64 and \citet{Landolt1983}'s stars 98-978 and 102-1081. 
An in-depth analysis of these calibration stars is provided in \citet{Marsset2020}. 

Data reduction and spectral extraction followed the procedure outlined in \citet{Binzel2019}. 
We summarize it briefly here. 
Reduction of the spectral images was performed with the Image Reduction and Analysis Facility (IRAF) \citep{Tody1993} and Interactive Data Language (IDL), using the Autospex software tool to automatically write sets of command files \citep{Rivkin2005}. 
Reduction steps for the science target and its corresponding calibration stars included trimming the images, creating a bad pixel map, flat fielding the images, sky subtracting between AB image pairs, tracing the spectra in both the wavelength and spatial dimensions, co-adding the spectral images, extracting the 2-D spectra, performing wavelength calibration, and correcting for air mass differences between the asteroid and the solar analogs. 
Finally, the resulting asteroid spectrum was divided by the mean stellar spectra to remove the solar gradient.

\section{spectroscopic results} \label{sec:result}
We compared the spectra of 596 Scheila obtained before and after the 2010 impact to search for any spectral change induced by this event. 
Spectral data obtained before the impact event were retrieved from \citet{Bus2002}, \citet{DeMeo2009}, and \citet{Licandro2011EPSC}.
NIR spectra collected after the impact event are from \citet{Yang2011} and \citet{Licandro2011EPSC}, as well as from this study.
Table \ref{tab:1} shows the sub-observer coordinates of the asteroid during the pre- and post- impact NIR observations calculated based on the shape model from \citet{Hanus2021}.
Since a visible reflectance spectrum is not available after the impact, the average of the multicolor photometry of \citet{Trigo-Rodriguez2011EPSC}, \citet{Betzler2012}, \citet{Hsieh2012}, \citet{Jewitt2012}, and \citet{Shevchenko2016} was utilized instead.

\begin{deluxetable*}{llcccccl}
\label{tab:1}
\tablenum{1}
\tablecaption{List of  the sub-observer coordinates of 596 Scheila at the time of NIR observations}
\tablewidth{0pt}
\tablehead{
\colhead{Observation time}&\colhead{Timing of }&\multicolumn4c{Sub-Earth point}&\colhead{Spectral slope}&\colhead{Reference}\\
\colhead{}&\colhead{Observation}&\colhead{${\lambda_{1}}^{a}$ [\degr]}&\colhead{${\phi_{1}}^{a}$ [\degr]}&\colhead{${\lambda_{1}}^{b}$ [\degr]}&\colhead{${\phi_{1}}^{b}$ [\degr]}&\colhead{[\% $\mu$m$^{-1}$]}&\colhead{}
}
\startdata
2002/06/01 09:40&before impact&137&46&232&$-$47&25.5&\cite{DeMeo2009}\\ 
2010/12/10 10:58&after impact&157&$-$40&292&41&$36.5^{c}$&\cite{Licandro2011EPSC}\\ 
2010/12/10 14:35&after impact&75&$-$40&210&41&$36.5^{c}$&\cite{Licandro2011EPSC}\\ 
2010/12/10 15:02&after impact&64&$-$40&200&41&$36.5^{c}$&\cite{Licandro2011EPSC}\\ 
2011/01/04 14:50&after impact&132 &$-$39&263&43&$40.5^{d}$&\cite{Yang2011}\\
2011/01/05 14:20&after impact&319 &$-$39&90&44&$40.5^{d}$&\cite{Yang2011}\\
2011/01/07 02:56&after impact&209&$-$39&339&44&$36.5^{c}$&\cite{Licandro2011EPSC}\\ 
2011/01/08 06:22&after impact&306&$-$39&76&44&$36.5^{c}$&\cite{Licandro2011EPSC}\\ 
2011/01/09 06:50&after impact&110&$-$39&241&44&$36.5^{c}$&\cite{Licandro2011EPSC}\\ 
2011/01/10 01:53&after impact&39&$-$39 &168&44&$36.5^{c}$&\cite{Licandro2011EPSC}\\ 
2011/01/10 07:02&after impact&282&$-$39&51&44&$36.5^{c}$&\cite{Licandro2011EPSC}\\ 
2011/02/07 09:31&after impact&99&$-$40&225&50&46.7&This study\\ 
\enddata
\tablecomments{
\\ 
There are two shape/pole solutions, which is a common thing in the inversion of disk-integrated data without stellar occultations or disk-resolved data.\\ 
$^{a}$Pole solution 1: ${\lambda}_1$ = 110 [\degr], ${\beta}_1$ = $-$24 [\degr], Period = 15.8605 [hour]\\ 
$^{b}$Pole solution 2: ${\lambda}_1$ = 273  [\degr], ${\beta}_1$ = $-$38  [\degr], Period = 15.8609 [hour]\\
$^{c}$The value of the combined spectrum in \citet{Licandro2011EPSC}.\\
$^{d}$The value of the combined spectrum in \citet{Yang2011}.\\
}
\end{deluxetable*}

\begin{figure*}
\gridline{\fig{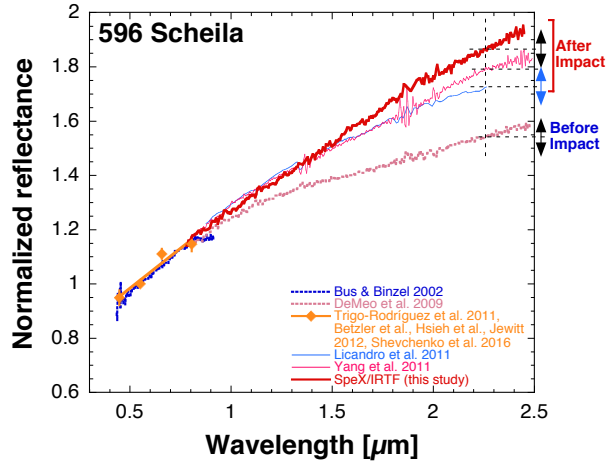}{0.5\textwidth}{}
          }
\caption{
Comparison of 596's NIR spectra before and after the 2010 impact event.
The dotted and solid lines show the spectra before and after the impact event, respectively.
The figure shows spectra of 596 in the 0.4--2.5 $\mu$m wavelength region, normalized to 1 at 0.55 $\mu$m.
The black and blue arrows in the figure indicate the slope uncertainty from \citet{Marsset2020}, and the dispersion range of the spectra obtained by \citet{Licandro2011EPSC}, centered at the reflectance value of the spectra at 2.35 $\mu$m as shown by the black dashed lines.
}
\label{fig:spectra}
\end{figure*}

\citet{Marsset2020} estimated the uncertainty of spectral slope measurements of asteroids in the NIR wavelength region from 20 years of MITHNEOS observations with the SpeX spectrograph on the IRTF, and showed that the uncertainty is 4.2 \% $\mu$m$^{-1}$.
Spectra obtained by \citet{Yang2011} after the impact event, which are not from MITHNEOS observations, and the one obtained in this study, which is from MITHNEOS observations, are consistent within that uncertainty range (Figure \ref{fig:spectra}).
The total exposure time of the two spectra acquired in \citet{Yang2011} is 400 s each, and the total exposure time of the pre- and post-impact spectra acquired by MITHNEOS observations is 840 s and 1800 s (\citealt{DeMeo2009} and this study), respectively: these spectra are data of the comparable quality with signal-to-noise ratios of several hundreds.
In addition, the spectra collected after the impact event by \citet{Licandro2011EPSC} show that the variance of the slope is about 10\%, while the spectrum obtained by \citet{Yang2011} is within this range  (Figure \ref{fig:spectra}).
Considering the dispersion and uncertainty mentioned above, the agreement of the spectra after the impact event is consistent with the point made by \citet{Licandro2011EPSC} that the NIR spectra of 596 does not change with rotation phase.
The spectrum of \citet{Yang2011} in Figure \ref{fig:spectra} is a combination of two days of data, and the difference between the two data sets is several percent.
This is also an indication of the homogeneity of the surface of the 596 after the impact event.

Comparing the pre-impact spectrum of \citet{DeMeo2009} obtained by MITHNEOS with the post-impact spectrum of this study also obtained by MITHNEOS shows that the difference is significantely larger than the indeterminacy obtained by \citet{Marsset2020}, which indicates that the spectrum changed before and after the impact event  (Figure \ref{fig:spectra}).
The slope of 596 at 0.8--2.5 $\mu$m before and after the impact event is 25.5 and 46.7 \% $\mu$m$^{-1}$, respectively.
In fact, the spectral type of 596 before the impact event was T-type \citep{DeMeo2009}, but after the impact event it changed to D-type \citep{Yang2011}.
This change is unlikely to be due to varying observing conditions. 
In particular, the two IRTF spectra obtained as part of MITHNEOS before and after the impact were taken at slightly different airmass values (1.4 before the impact versus 1.1 after). 
\citet{Marsset2020} showed that air mass only contributes to a systematic slope change of $-$0.92 \% $\mu$m$^{-1}$ per 0.1 unit air mass: this effect is insufficient to explain the slope difference between the two spectra.

Possible additional external factors causing slope variability of the asteroid may include varying weather conditions and close encounters of the asteroid with a bright star during the observations.
However, the sky was reported to be clean during the night of the pre-impact observations and no stellar encounter was confirmed based on a combination of the JPL HORIZONS ephemeris generator\footnote{https://ssd.jpl.nasa.gov/horizons/} and the SIMBAD astronomical Database\footnote{http://simbad.u-strasbg.fr/simbad/\label{SIMBAD}} . 
In addition, a search for bright (V-band magnitude  \textless\ 10.5) stars using the SIMBAD Astronomical Database\textsuperscript{\ref{SIMBAD}} reveals no such stars at proximity (within 20\arcmin) of 596 during the MITHNEOS observations. 
Finally, another asteroid observed on the same, pre-impact night as 596, the S-type asteroid 57 Mnemosyne, was also observed on 2010-07-11 and exhibited an entirely consistent shape of its spectra between the two observing nights (see Figure \ref{fig:57}).

\begin{figure*}
\gridline{\fig{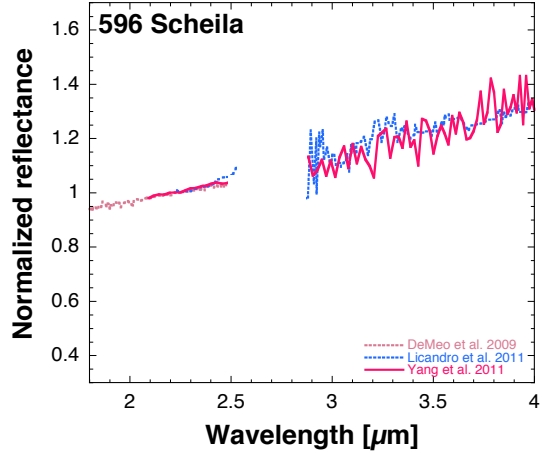}{0.5\textwidth}{}
          }
\caption{
Comparison of the 3-$\mu$m spectra of 596 before and after the 2010 impact event.
The dotted and solid lines show the spectra before and after the impact event, respectively.
The figure shows the spectra in the 1.8--4.0 $\mu$m wavelength regions, normalized to 1 at 2.2 $\mu$m
}
\label{fig:spectra3}
\end{figure*}

Based on the change in the lightcurve of 596 recorded in \textit{R}-band filter before and after the impact event \citep{Ishiguro2011a}, there may be an increase or decrease in visible reflectance of up to 4.5\%.
However, the slopes of the visible spectra taken before and after the impact event are consistent within the observational error range, and it can be said that there is no significant change in the visible slope before and after the impact event.
Spectral observations of 596 in the 3-$\mu$m band were also made before and after the impact event, and it can be said that the 3-$\mu$m spectra are consistent within the observational error range (Figure \ref{fig:spectra3}).
\citet{Yang2011} found that there was almost no ice on the surface of 596 through 3-$\mu$m spectra of 596 after the impact event.
Similarly, the 3-$\mu$m spectrum of 596 obtained before the impact \citep{Licandro2011EPSC} does not exhibit any water ice signature either. 

\section{Discussion} \label{sec:Discussion}
The slope of the NIR spectrum of 596 Scheila after the impact, with 46.7 $\mu$m$^{-1}$, is redder than the spectrum before the impact event,  with 25.5 \% $\mu$m$^{-1}$.
This section discusses the factors that may have contributed to this.

The first possibility is surface heterogeneity.
Table \ref{tab:1} shows the sub-observer coordinates of the pre- and post-impact observations for two different solutions of the pole orientation. 
Since 596 before the impact was observed in June, and 596 after the impact was observed around January, the direction of observation of 596 from the Earth is different by about 180 degrees.
Sub-Earth latitude are reversed because the observation period before and after the impact happens to be off by half a year.
The homogeneity of the post-impact spectra indicates that the southern hemisphere\footnote{This applies to the first shape/pole solution. For the second shape/pole solution, the northern and southern hemispheres are reversed.\label{ASAS-SN}}, observed after the impact, exhibits an homogeneous surface. 
Comparing the surface of the southern hemisphere\textsuperscript{\ref{ASAS-SN}} observed after the impact and the surface of the northern hemisphere\textsuperscript{\ref{ASAS-SN}} observed before the impact may be heterotropic.
However, considering that most of the material from the crater was expected to be ejected isotropically \citep[e.g.,][]{Schultz1985,Shuvalov2011}, and that the ejected dust covered the entire surface of 596 as it became comet-like \citep[e.g.,][]{Ishiguro2011a,Ishiguro2011b}, we expect no difference between the northern and southern hemisphere spectra after the impact.

Note that the argument for partial resurfacing was based on the presence or absence of changes in the visible lightcurve before and after the impact event \citep{Bodewits2014}.
However, the present study does not show any change in the visible spectrum before and after the impact, and the maximum albedo change is 4.5\% \citep{Ishiguro2011a}.
NIR spectra after the impact were obtained at longitudes with both many and little changes in the visible lightcurve before and after the impact event, but they show a similar spectral slope.
Therefore, it is appropriate to assume that the spectral change occurred in the whole area observed by NIR wavelength region before and after the impact event.
These results indicate that the only-partial resurfacing of 596 inferred from visible lightcurve may be an underestimation of the true ejecta distribution.
Since this study shows that the NIR region is more sensitive to spectral changes before and after the impact event than the visible wavelength region, it is natural to assume that ejecta were deposited over the entire surface of 596.

\begin{figure*}
\gridline{\fig{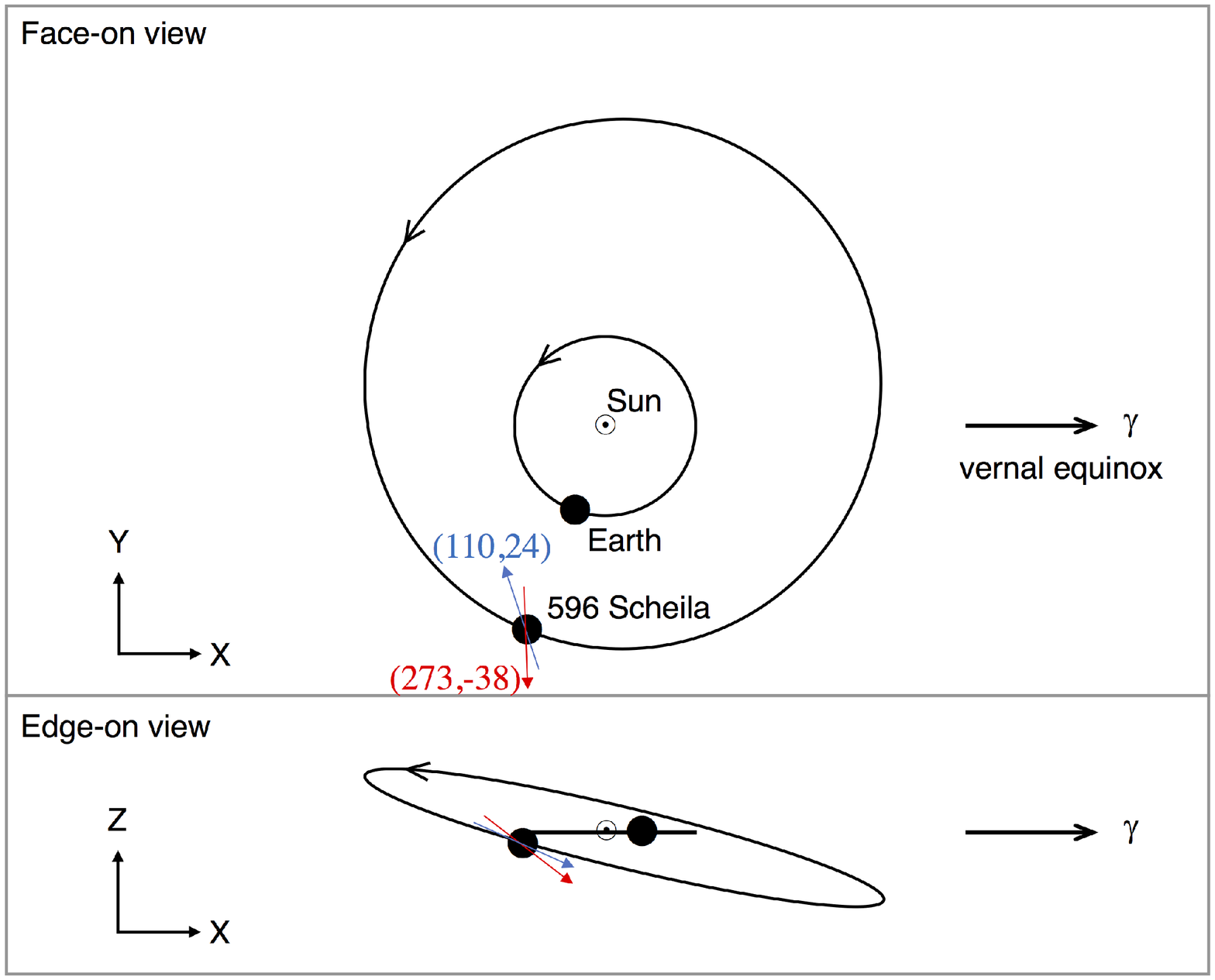}{0.5\textwidth}{}
          \fig{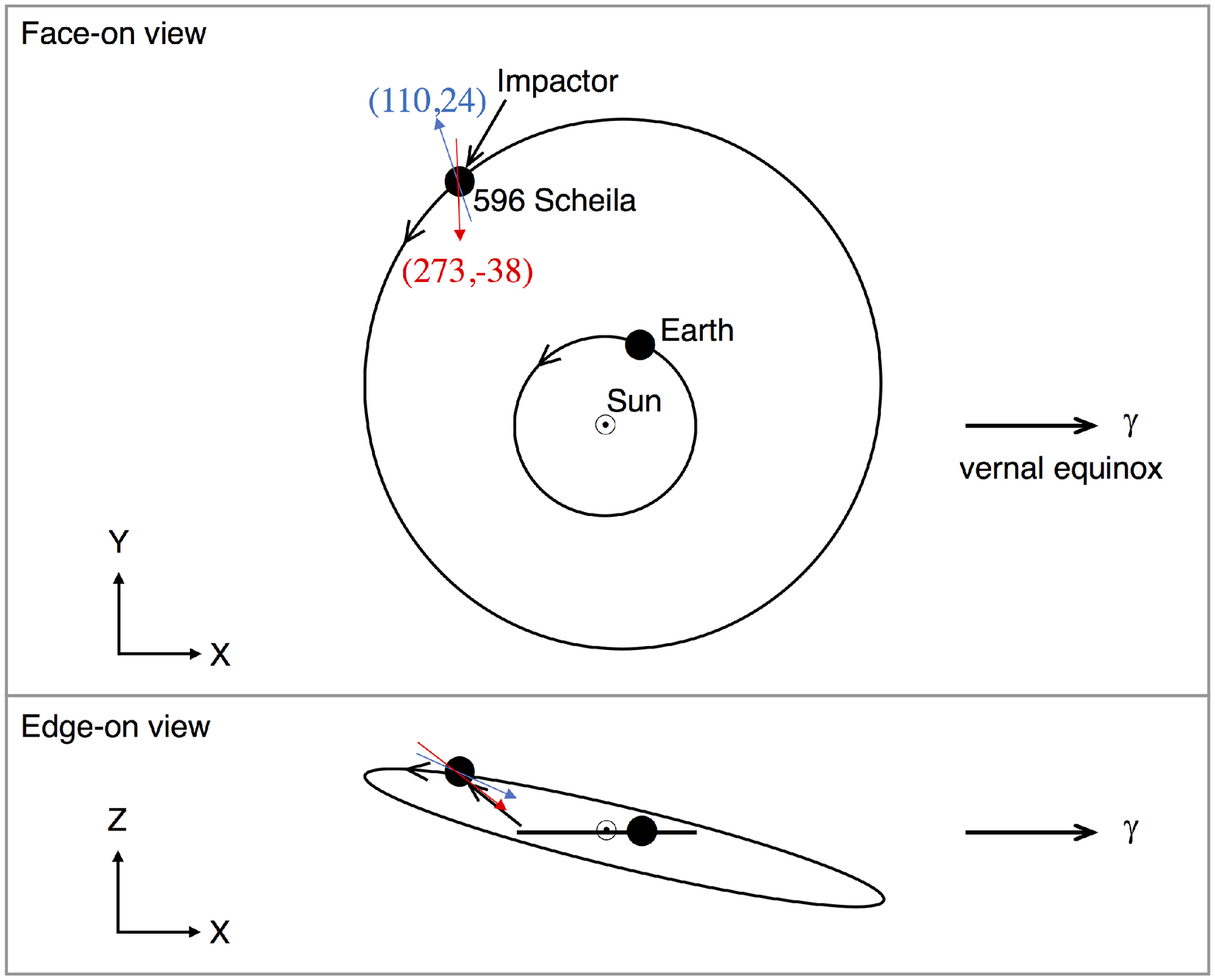}{0.5\textwidth}{}
          }
\caption{Orbit diagrams of 596 at the time of pre-impact NIR observation (left panel) and the 2010 impact event (right panel).
Assuming an impact angle of 45\degr, which is the highest impact probability, the small asteroid would have collided with 596 from the direction ($\alpha$, $\delta$) = (60\degr, -40\degr) \citep{Ishiguro2011b}.
The values in blue and red fonts are the values of the two pole solutions of 596 \citep{Hanus2021}.
}
\label{fig:impact}
\end{figure*}

Also, the area where the impactor is thought to have collided with 596 \citep{Ishiguro2011b} is at approximately the same latitude as the one observed before the impact event (Figure \ref{fig:impact}).
This shows that the surface material in the area observed before the impact must have fallen the area observed after the impact event as ejecta.
In the case of spectral heterogeneity between the northern and southern hemispheres of 596, the spectra should have been similar to the spectrum taken before the impact, however, the pre- and post-impact spectra are different, indicating that geographic heterogeneity is not responsible for the observed spectral change of 596.
%

The second possibility is that the post-impact spectrum is redder due to a change in size of the particles that compose the surface layer of 596.
The reddening of the spectrum after the impact suggests that finer particles may be present compared to before the impact \citep[e.g.,][]{Johnson1973,Vernazza2016}.
Fine particles ejected from craters by the excavation flow can have a wide range of ejection velocities \citep{Singer2020}. 
However, finer particles have faster ejection velocities on average than larger particles, meaning that large particles should have preferentially been re-accreted at the surface of 596 after the impact.
In general, if large size particles remain, the spectrum should change to blue, but the observed change, a spectral reddening, is the opposite (see also Figure \ref{fig:Tagish}).
Note that the original surface layer of 596 must also be composed of particles ejected from previous impact events, so the particle size distribution should basically be the same before and after the impact event.
Due to thermal fatigue  and meteoroid impacts \citep[e.g.,][]{Delbo2014,Basilevsky2015}, the surface particles of asteroids tend to become smaller as time goes by after large impacts such as the 2010 impact.
Given that, it is not possible to explain spectral reddening by grain size effect.

The third possibility is that the spectral slope change for the case of 596 may be due to the presence of the dust coma.
The large dust particles that are inside the Hill radius of 596 and are not affected by radiation pressure may be around 596 for a long time.
As shown in \citet{Ishiguro2011b}, dust more than 140 $\mu$m may be mixed with the spectrum of 596 for a long time (Dust less than 140 $\mu$m is stretched by radiation pressure  on 2011 February, when the observations for this study were made.). 
However, since after the impact about a few weeks the visible apparent magnitude of 596 approached the predicted value from before the impact \citep[e.g.,][]{Jewitt2011,Trigo-Rodriguez2011EPSC}, it is natural to assume that the scattering cross section of such a dust cloud is negligible compared to the entire surface of 596.
Also, if the dust coma was affecting the spectrum, the spectrum from the observation closest to the collision event may be the reddest, but Figure \ref{fig:spectra} and \citet{Licandro2011EPSC} show that this is not the case.

Finally, space weathering is another mechanism that may have also played a role in the spectral change of 596 during the impact.
The outer layer of asteroids is thought to undergo space weathering, where the spectrum is altered by micrometeorite impacts and solar wind irradiation \citep[e.g.,][]{Clark2002,Brunetto2015}.
Most of the particles that cause cometary activity due to impacts or ice sublimation are thought to be ejected from the asteroid's subsurface rather than from the surface (\citealp[e.g., impacts:][]{Stoeffler1975}; \citealp[ice sublimation:][]{AHearn2005}).
If covered by those particles, the surface layer after cometary activity would be by a fresh material that has not undergone space weathering.

Experiments simulating space weathering have been conducted on the C2-ungrouped Tagish Lake meteorite, which has a similar spectrum to that of 596. 
In these experiments, the spectrum tends to become bluer after space weathering\footnote{The spectral slope of dark carbonaceous chondrites (CI and CM) including the Tagish lake meteorite becomes bluer due to space weathering, while that of bright carbonaceous chondrites (CO and CV) becomes redder.} \citep{Hiroi2013LPSC,Vernazza2013,Lantz2017}.
Along these lines, \cite{Lantz2018} also stated that Tagish Lake-like parent bodies are hidden in X- and T-type objects based on the results of laboratory space weathering simulations.
Since the spectrum of 596 before the 2010 impact event is  bluer (T-type) \citep{Bus2002,DeMeo2009}, the fact that the surface of 596 before the impact was space weathered is in agreement with the claim of \cite{Lantz2018}.

The maximum increase or decrease in absolute reflectance of 596 after the impact in the \textit{R}-band is about 4.5\% \citep{Ishiguro2011a}, which is very small compared to the spectral change obtained at minimum irradiance in the space weathering experiment of \citet{Lantz2017}.
At minimum irradiance, the visible spectrum does not change at 0.4--0.9 $\mu$m wavelength region (see the yellow area in Figure \ref{fig:Tagish}), and becomes bluer at 0.9--2.5 $\mu$m wavelength region (see the blue area in Figure \ref{fig:Tagish}). 
This trend in this laboratory experiment is highly consistent with the spectral change of 596 in the same wavelength range before and after the impact event.
Note that the lightcurves acquired in \textit{R}-band were measured as relative values, and it is not possible to determine whether the absolute reflectance has increased or decreased after the impact only by comparing the lightcurves collected pre- and post-impact.

Previous examples of observations of fresh surfaces for small solar system bodies include: 1: P/2010 A2, which was shattered by a catastrophic disruption \citep{Kim2012,Kim2017}; 2: 6478 Gault, which is believed to have blown away its surface dust due to increased rotation \citep{Marsset2019,Purdum2021}; and 3: 162173 Ryugu, on the surface of which an artificial crater was formed by the  Small Carry-on Impactor of Hayabusa2 spacecraft \citep{Arakawa2020,Morota2020}. 
Due to the lack of spectroscopic observations acquired before the destruction of P/2010 A2, it is not known whether the spectrum of that object changed as a consequence of its disruption. 
In the case of 6478, important spectral bluing occurred and was attributed to the loss of surface regolith \citep{Marsset2019}. 
Finally, the artificial crater experiment on the surface of the C-type asteroid Ryugu resulted into a slightly bluer visible light spectrum of that region of the asteroid \citep{Morota2020}. 
The different spectral trend observed for these objects compared to 596 is not entirely surprising considering their distinct spectral type (hence surface composition), environment (heliocentric distance, hence surface and subsurface thermal evolution), and resurfacing mechanisms.
Our results provide the first spectroscopic evidence that the space-weathered surface layer of an asteroid was covered by fresh material from the subsurface during a natural impact event.
Meanwhile, this is also the first time that actual astronomical observations show that D-type asteroids with spectra similar to the Tagish Lake meteorite become less red as a consequence of space weathering.
Our results are also the first to provide astronomical evidence that Tagish-Lake parent bodies may be found among asteroids that are spectrally bluer than that meteorite, a prediction previously made by \citet{Lantz2018} based on laboratory experiments.

It is generally accepted that the Tagish Lake meteorite, which has been collected as a meteorite with a size large enough to conduct space weathering experiments, is of D-type asteroid origin.
\citet{Vernazza2015} stated that most surfaces of dark C- and X-complex and D-type (including T-type) asteroids have not been collected as meteorites on Earth, but only as anhydrous chondritic porous interplanetary dust particles (IDPs).
\citet{Hasegawa2017} also pointed out that in addition to their spectroscopic properties, the radar and polarization properties of these bodies indicate that they are related to chondritic porous IDPs.
Most dark X-complex and D-type asteroid surfaces have been shown to have anhydrous surface compositions by spectroscopic observations in the 3-$\mu$m wavelength region \citep[e.g.,][]{Takir2012,Usui2019}, and 596 also has anhydrous surface layers, which indicates an association with chondritic porous IDPs (Figure \ref{fig:spectra3}).
Meanwhile, the Tagish Lake meteorite which is aqueously altered and exhibits a 3-$\mu$m absorption feature \citep{Hiroi2001}, meaning that it may not be exactly analog in composition to 596.
\citet{Vernazza2021} showed that large dark X-complex and D-type asteroids that were implanted inward of 3.3 au would have lost outer shell due to water-ice sublimation and impact erosion. 
However, since the erosion rate at the location of 596 is only 1 km over the age of the solar system \citep{Vernazza2021}, it is thought that 596 does not expose any of aqueously altered materials expected to be contained and abundant in its interior \citep{Vernazza2017a}.

Our finding of a collision-induced change in the spectral slope and the spectral type of 596 supports the idea that more D types were originally implanted in the main belt than previously thought.
This agrees with \citet{Vernazza2021}'s proposal that dark X-complex and D-type and some C-type asteroids may share the same origin in the trans-planetary and/or trans-Neptunian planetesimal region.
Most of the heavily weathered D-type asteroids may be very red objects, such as those discovered by \citet{Hasegawa2021b}, that have become bluer in their spectra as a result of space weathering.
Our discovery therefore relax the observational constraints on dynamical models simulating the early implantation of outer solar system bodies in the asteroid belt. 
For instance, \citet{Levison2009} used the low observed number of dark X and D-type asteroids as a stringent constraint to their simulations. 
Our results that some implanted objects may in fact be hidden among moderately red asteroids loosens that constraint.

Assuming that a collision event of the magnitude of the 2010 impact on 596 would entirely refresh the surface of an asteroid, the probability of an asteroid collision at 596's location in the main-belt, the number of asteroids with similar size to the impactor of 596, and the cross-sectional area of 596 yields a resurfacing timescale of 10$^{3.5}$ -- 10$^{4.5}$ years \citep{Jewitt2012}.
Although it is not clear how much smaller an impact can be while still being able to resurface 596, this means that the typical timescale for surface renewal is shorter than 10$^{3.5}$ -- 10$^{4.5}$ years.

\section{Conclusions} \label{sec:Conclusions}
We compared the spectra of 596 Scheila obtained before and after the 2010 impact event on that asteroid and showed that the NIR spectrum of its post-impact fresh surface was redder than the NIR spectrum of its space-weathered surface before the impact.
This is consistent with the results of experiments simulating space weathering of the Tagish Lake meteorite, which has a spectrum similar to that of 596 after the impact.
Our results are the first spectroscopic evidence of asteroid resurfacing by fresh material from its subsurface during a naturally occurring impact, and the first observational evidence that the spectra of D-type asteroids become less red owing to space weathering.
Our results also imply that under space weathering process, carbon-free, silicate-rich assemblages will redden whereas carbon-rich assemblages will be bluish.
Finally, our results suggest that the number of implanted outer Solar System objects in the main-belt is much higher than previously estimated, but many of these objects are hidden below a space weathered crust. 

\appendix
\section{Spectra of 57 Mnemosyne} \label{sec:57}
\begin{figure*}
\gridline{\fig{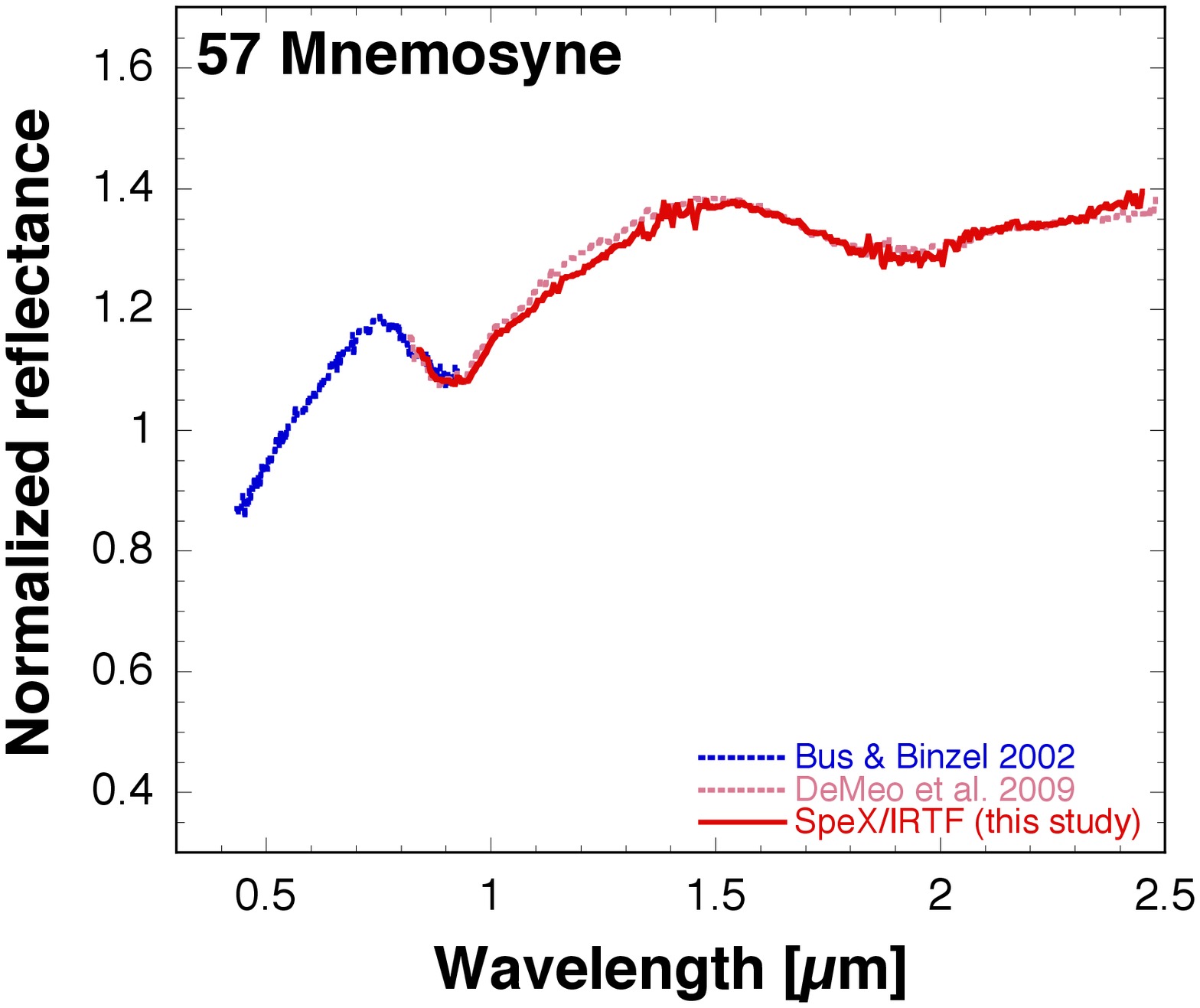}{0.5\textwidth}{}
          }
\caption{
Spectra of 57 Mnemosyne.
}
\label{fig:57}
\end{figure*}

The NIR spectrum of 57 Mnemosyne was obtained on the same day that the spectrum of 596 was obtained before the impact event \citep{DeMeo2009}.
The spectra were acquired on different days, and both spectra are consistent (Figure \ref{fig:57}).
This shows the homogeneity of surface of 57, and also indicate that the observing conditions were stable and good on the day the two spectra were acquired.

\restartappendixnumbering

\section{Spectra of the Tagish lake meteorite} \label{sec:Tagish}
\begin{figure*}
\gridline{\fig{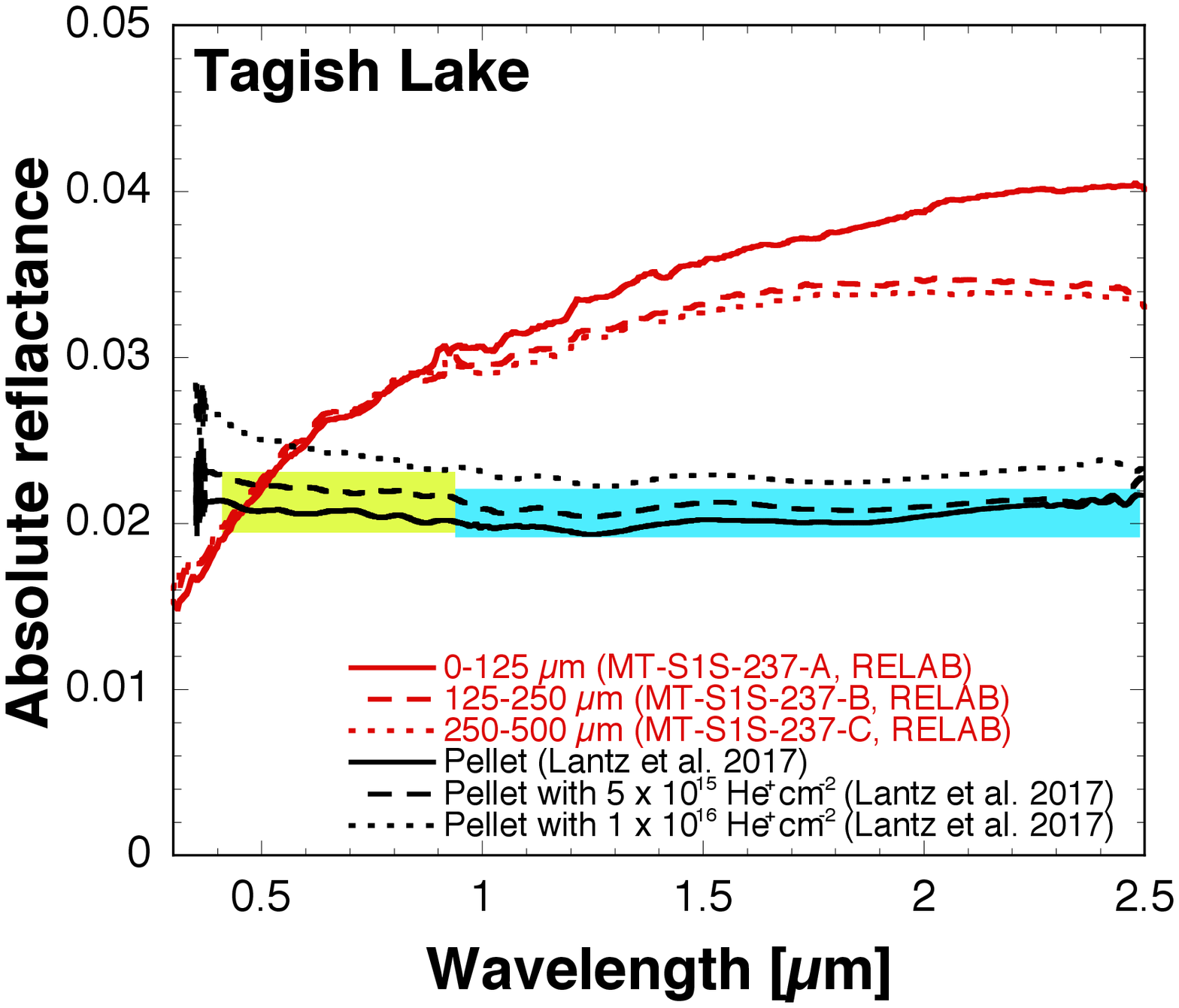}{0.5\textwidth}{}
          }
\caption{
Spectra of the Tagish Lake meteorite.
Yellow and cyan areas indicate regions where the slope of the spectrum did not change and regions where it did change after minimum irradiation, respectively.
}
\label{fig:Tagish}
\end{figure*}

Spectroscopic data of the Tagish Lake meteorite is provided for reference.
Powder data and pellet data of the Tagish Lake meteorite were obtained from the Reflectance Experiment Laboratory (RELAB)\footnote{http://www.planetary.brown.edu/relab/} at Brown University and \citet{Lantz2017}, respectively  (Figure \ref{fig:Tagish}).
Pellet samples exhibit a flat spectral slope, and are likely not representative of asteroid regolith.
In general, the spectra of pellet samples are much bluer than those of powder samples \citep[e.g.,][]{Hasegawa2019}.

\begin{acknowledgments}
We would like to thank the referee for their careful review and constructive suggestions, which helped us to improve the manuscript significantly.  
We are grateful to Dr. Cateline Lantz for sharing her Tagish Lake spectral data. 
We would like to express our gratitude to Dr. Takahiro Hiroi for useful comments about the Tagish Lake data that he acquired at RELAB.
We greatly appreciate Dr. Masahiko Arakawa and Dr. Tachuhiro Michikami for useful comments about impact.
This work is based on observations collected at the Infrared Telescope Facility, which is operated by the University of Hawaii under contract 80HQTR19D0030 with the National Aeronautics and Space Administration. 
The authors acknowledge the sacred nature of Mauna kea and appreciate the opportunity to observe from the mountain. 
The spectra of Tagish Lake meteorite used in this study are from the RELAB database of Brown University.
This study has utilized the SIMBAD database, operated at CDS, Strasbourg, France, and the JPL HORIZONS ephemeris generator system, operated at JPL, Pasadena, USA. 
M.M. and F.D. were supported by the National Aeronautics and Space Administration under grant No. 80NSSC18K0849 and 80NSSC18K1004 issued through the Planetary Astronomy Program.
M.I. was supported by the NRF grant No. 2018R1D1A1A09084105.
The work of J.H. has been supported by the Czech Science Foundation through grant 20-08218S.
This study was supported by JSPS KAKENHI (grant nos. JP18K03723, JP19H00719, JP20K04055, JP21H01140, and JP21H01148) and by the Hypervelocity Impact Facility (former facility name: the Space Plasma Laboratory), ISAS, JAXA.
\end{acknowledgments}

%

\facilities{IRTF:3.0m}


\software{IRAF \citep{Tody1993}, IDL}



\bibliography{hase596}



\end{document}